\documentstyle[12pt]{article}
\def\la{\mathrel{\mathpalette\fun <}}

\def\fun#1#2{\lower3.6pt\vbox{\baselineskip0pt\lineskip.9pt
\ialign{$\mathsurround=0pt#1\hfil##\hfil$\crcr#2\crcr\sim\crcr}}}

\newcommand {\numu } {$\nu_{\mu}$}
\newcommand {\nuone } {$\nu_{1}$}
\newcommand {\nutwo } {$\nu_{2}$}
\newcommand {\nutau } {$\nu_{\tau}$}
\newcommand {\nue } {$\nu_{e}$}

\newcommand {\nuebar } {$\bar\nu_{e}$}

\begin{document}
\begin{flushright}
ULB-TH/96/14\\
hep-ph/9608266
\end{flushright}

\vskip1cm
\begin{center}
{\large{\bf Stimulated Neutrino Conversion and Bounds
\vskip0.5cm

on Neutrino
Magnetic Moments}}
\vskip1cm

{\bf J.-M. Fr\`ere} \footnote{postal address: Physique Theorique CP 225;
U.L.B. Boulevard du Triomphe; B-1050 Bruxelles, Belgium; email:
frere@ulb.ac.be}\\
Universit\'{e} Libre de Bruxelles \\
\vskip0.5cm

{\bf R.B. Nevzorov, M.I. Vysotsky}\\
ITEP, Moscow 117259, Russia

\vspace{1cm}

\noindent {\bf Abstract}
\end{center}

\vskip1cm

\noindent Recent experiment proposed to observe induced radiative
neutrino transitions
are confronted to existing bounds on neutrino magnetic moments from
earth-based experiments. These are found to exclude any observation
by several orders of magnitude,
unless the magnetic moments are assumed to be strongly
momentum dependent. This possibility is discussed in some generality,
and we find that nontrivial dependence of the neutrino form factor
may indeed occur, leading to quite unexpected effects, although
this is insufficient by orders of magnitude to justify the experiments.

\newpage
The following method of search for neutrino magnetic transitions was
recently proposed in \cite{1}: neutrinos pass a resonant cavity full of
low-energy photons which induce the neutrino conversion:
\nuone $\gamma \to$ \nutwo. While neutrino decay in
vacuum is strongly suppressed by lack of phase space
if the mass difference between
\nuone  and \nutwo  is small ($(\Gamma \sim (\delta m)^3)$)
in the case of stimulated conversion
the energy is provided by the electromagnetic field, and the  $\delta m$
suppression disappears. This fact makes a suggested way to look for
the neutrino transitional magnetic moments very attractive. In
paper \cite{1} possible experiments with the electron neutrinos from
a reactor or the Sun were discussed. The same idea was applied in
paper \cite{2} to accelerator-produced \numu 's.
The scheme of
the experiment is obvious: neutrinos on the way to a standard detector
pass through a high-quality resonant
cavity;   a variation in
the detected neutrino fluxes associated to switching on and off the
cavity is interpreted as the signal of a radiative transition.
According to the estimates given \cite{2}, the suggested experiments
could hope to detect
transition neutrino magnetic moments $\mu$ larger than:
\begin{equation}
\mu \geq  4 \cdot 10^{-4} \mu_B\;\;,
\label{1}
\end{equation}
where $\mu_B$ is Bohr magneton.
\vskip .5cm

Our purpose here is to study the compatibility of such large
transition magnetic moments with existing bounds from earth-based
experiments.
A positive signal of stimulated transition could be observed
in three different situations: (in what follows, we only mention neutrinos,
parallel situations obviously happen with their antiparticles, or rather,
their CP conjugates):

- apparition : an inactive (sterile) neutrino present in the beam
is converted inside the cavity into an active one.
The initial neutrino can be either a right-handed component of
one of the known neutrinos, or a new, this far undetected particle
altogether, while the neutrino emerging from the cavity is one of
the known, (mostly) left-handed types (\nue, \numu, \nutau  or
 there antiparticles)

- disparition : an active neutrino disappears into an inactive species,
either a new state, or the right-handed component of a (Dirac) known
neutrino

- conversion : a known neutrino is converted into another species, itself
observable , for instance a left-handed \numu  absorbs a $\gamma$ and turns
into a right-handed (active) \nuebar; the beam then appears depleted in one
species, and enriched in another.

What is common to all theses possibilities is that they always involve at
least one transition moment to or from one of the 3 standard, relatively
well-known neutrinos.

The bounds on neutrino magnetic moments are usually quoted in the
litterature in terms of "diagonal" moments rather than transition
moments. They are\cite{3}:
\begin{equation}
\mu_{\nu_{\mu}} \leq 10^{-9} \mu_B\;\;, \mu_{\nu_e} \leq 10^{-10}
\mu_B\;\;.
\label{2}
\end{equation}
The bound on the  tau neutrino magnetic moment is stronger than
(\ref{1}) as well  \cite{3}:
\begin{equation}
\mu_{\nu_{\tau}} \leq 5.4 \cdot 10^{-7} \mu_B\;\;.
\label{3}
\end{equation}

We want to point out that the above bounds are in fact more general
and do apply as well to transition magnetic moments, provided phase space
is sufficient in the experiment considered.
For this, it is easiest to reason in terms of helicity amplitudes,
a natural approach for light, energetic neutrinos.
Interference between helicity amplitudes only occurs through the
neutrino masses, and can thus be safely neglected for most purposes.

These bounds above are deduced from the absence of any observed addition
to the predicted standard model neutrino-electron
scattering.
While standard model weak contributions, mediated by the Z and W direct
couplings
are diagonal in the helicity representation (LL and RR transitions), as
expected from a regular gauge interaction, an hypothetical magnetic coupling
to the photon, being a tensor interaction, involves a LR transition.
For this reason, the neutrino magnetic moment contribution never
interferes with the standard contribution, and the cross sections thus
add trivially.
As a result, the respective natures of the initial and final neutrino
are irrelevant, provided the mass of the final neutrino does not suppress
the reaction.
Thus, the resulting cross section (and the resulting bound) is the same,
whether the final (undetected) neutrino is identical to (diagonal magnetic
moment) or different from the first (transition moment).
The bounds (\ref{2}), (\ref{3}) are
applicable to any kind of a transitional magnetic moment, the only
condition being the smallness of the mass difference between the two
neutrino species, compared to the center of mass energy
available in the process, which is about 100 MeV for $\nu_{\tau}$
and several MeV for $\nu_e$  and $\nu_{\mu}$.

The comparison of (\ref{1}), (\ref{2}), (\ref{3})
then seems to settle the issue.
\vskip .5cm

A potential loophole however exits.
The bounds (\ref{2}), (\ref{3}) were indeed obtained for virtual
photons, with typical
$q^2$ varying from $\sim(100 {\rm
MeV})^2$ for $\nu_{\tau} e$-scattering and $\sim (1 {\rm MeV})^2$ for
$\bar{\nu}_e e$ scattering,
while radiative decays of neutrinos deals with real photons and
absorption of a photon from the resonance cavity involves
$q^2 \sim(10^{-6} {\rm eV})^2$.
So we should, as devil advocates, consider the possibility that
neutrino magnetic
form factors build up with diminishing $|q^2|$, in which case even
a strong upper bound
at large $q^2$ (\ref{2}),(\ref{3}) could coexist with a considerably
larger $\mu$ at small $q^2$, restoring hope for a
cavity experiment.
Let us look more closely at this possibility.

The neutrino magnetic moment in generalizations of the standard model
is usually generated through a triangle diagram involving
virtual charged particles.  The dependence on $q^2$ is determined  by
the mass of the heaviest charged particle in the loop (usually it is
a new scalar or vector boson) and for $q^2 < M^2$ has the following
form:
\begin{equation}
\mu(q^2) = \mu(0)[1 + a\frac{q^2}{M^2}]\;\;.
\label{4}
\end{equation}
Here $a$ is a model-dependent number of order 1,
while we know that $M$ has to be larger than 60 GeV
from the absence of new charged particles at LEP 1.5.
It is thus clear from (\ref{4}) that
$\mu((100 {\rm MeV})^2) = \mu(0)$ with very high accuracy.
Of course, renormalisability of the theory tells us that the formfactor
has ultimately to decrease with $q^2$ for asymptotic  $q^2$, but
this only really takes place for 
$q^2 > M^2$; and as there are no charged bosons
lighter than 60 GeV this cannot help us.

At this point, we might again give up, but devil's advocates tend
to perseverate, and another possibility exists.
The problem in generating the magnetic moment through charged particles
loops, is that the masses of the new intermediaries have to be very
large, as we know form LEP limits.
The situation is different with neutral intermediaries, but the 
counterpart is that their coupling to photons can only be through an 
induced magnetic moment. 
We will consider as an example the situation where the transition
magnetic moment between \nuone  and \nutwo \quad is induced through a loop 
containing the new light fermion $N$, and a light scalar $\varphi$; 
the light fermion is assumed to have a magnetic moment $\mu_N$,
possibly induced by heavy particles, and, basing ourselves on
the previous case, we will consider $\mu_N$ to 
be constant in most of the integration range (as we will see, the
infrared part of the integration is the relevant one).
It is quite clear that both $N$ and $\varphi$
should be SU(2) singlets to avoid conflict with
the measured $Z$-boson
invisible width.
For the simplicity of the argument, we will for now consider only
one neutrino and compute the induced diagonal magnetic moment;
the calculation below carries over essentially unchanged for transition
moments.

Let us take the standard neutrino \nuone  to be a Dirac particle 
coupling with $N$ through the Yukawa interaction $f \bar{\nu} N \varphi$,
to the new scalar particle. 
Let us note in passing that this interaction, if \nuone  is to be observable,
involves one element of an SU(2) doublet, with singlets. 
Such coupling is only possible as a consequence of the standard model
SU(2) breaking acting on an induced vertex;
we might therefore expect that it is somewhat suppressed, but this is 
not essential.
 
It is convenient to present the
neutrino magnetic formfactor in the following form:
\begin{equation}
\mu_{\nu} = \frac{f^2 \mu_{N}}{16\pi^2} I(m_N,
m_{\varphi}, m_{\nu}, q^2) \;\; ,
\label{5}
\end{equation}
where $I$ is a dimensionless function. For momentum transfer
$q^2$ larger than the masses of all three particles we get
a constant result $I= -1/2$ \footnote{this result is obtained by 
symmetrical integration in 4 dimensions. Other regularization methods
can differ by a constant term (e. g. dimensional regularization).
This however is typical of effective theories and 
does not affect our discussion},
instead of the anticipated decrease of the form factor (this is no real
surprise, as an implicit cut-off is present, and the 
decrease is only expected for momenta larger than the masses
of the charged particles responsible for the magnetic moment of the
fermion $N$). 
In most cases (except for fine tuning, see below) at $q^2 =0$ $I$ 
is of the order of unity, 
and the bounds on $\mu_{\nu}$ from the $\nu e$-scattering experiments
stay applicable to neutrino behavior
at small $q^2$, $\mu_{\nu}(q^2 =(100 {\rm MeV})^2)
 \approx \mu_{\nu} (q^2 =0)$.
However, some interesting exceptions
take place when the Feynman integral at small $q^2$ is infrared
divergent. For example, let us suppose that the masses of the external
neutrino and of the new spinor $N$ are almost degenerate, while the mass of
the scalar $\varphi$ almost vanishes (like in a Goldstone
boson situation):
\begin{equation}
 m_{\nu} \approx m_N \equiv m \;\; , \;\; m_{\varphi} \ll m_{\nu}
 -m_N \;\; .
\label{6}
\end{equation}
Performing the integration we get:
\begin{equation}
I = -\frac{1}{2} +\frac{4m^2}{-q^2}\left[ \ln (\frac{m^2}{m_N^2
-m_{\nu}^2}) -\frac{3}{4}\right]
\frac{2}{\sqrt{1+\frac{4m^2}{-q^2}}}
\ln \left( \frac{\sqrt{1+\frac{4m^2}{-q^2}}+1}{\sqrt{1+
\frac{4m^2}{-q^2}}-1}\right) +\delta(q^2) \;\; ,
\label{7}
\end{equation}
\begin{eqnarray}
\delta(q^2) & = & \frac{4m^2}{-q^2}\left[
\frac{1}{\sqrt{1+\frac{4m^2}{-q^2}}}
\ln \left( \frac{\sqrt{1+\frac{4m^2}{-q^2}}+1}{\sqrt{1+
\frac{4m^2}{-q^2}}-1}\right) \ln (4+\frac{-q^2}{m^2}) \right. +
\nonumber \\
& + & \left.\frac{2}{\sqrt{1+\frac{4m^2}{-q^2}}} 
\left( F(\frac{-1-\sqrt{1+\frac{4m^2}{-q^2}}}{2\sqrt{1+
\frac{4m^2}{-q^2}}}) -F(\frac{1-\sqrt{1+\frac{4m^2}{-q^2}}}{2\sqrt{1+
\frac{4m^2}{-q^2}}})\right)\right] \;\; ,
\label{81}
\end{eqnarray}
$$
F(\xi) =\int\limits^{\xi}_{0}\frac{\ln (1+x)}{x}dx \;\; ,
$$
\begin{equation}
I = -\frac{1}{2} \; , \;\; |q^2|\gg m^2
\label{91}
\end{equation}
\begin{equation}
 I = 4\ln (\frac{m}{2(m_N - m_{\nu})}) -\frac{7}{2} \; , \;\;
 |q^2| \la m^2
\label{10}
\end{equation}
In this case, the magnetic formfactor of neutrino at $q^2=0$ can
thus in principle be much larger
than at large momentum transfer.

Let us note that while the constant term in $I$ may depend on the
particular mehanism of $\mu_N$ generation at high energies, the
logarithmic term in $I$ originates in the infrared and is universal.

From (\ref{10}) and (\ref{5}), and the requirement to stay
in a regime where we can still trust perturbation theory,
 we can deduce the largest possible
value for the neutrino magnetic moment at $q^2 = 0$: 
(further tuning would at least force the resummation
of the leading terms)
\begin{equation}
\mu_{\nu}^{max}(q^2 =0) = \mu_N
\label{8}
\end{equation}

Even accepting such a strong tuning (and generalizing straightforwardly
to transition moments of nearly degenerate neutrinos) in a rather ad-hoc
scheme
does not allow any induced radiative transition to be observed in the suggested
cavity experiment, as strict bounds still apply to $\mu_N$.

We review them briefly, starting from earth-based experiments
and moving to astrophysical ones.

The bound from ref. \cite{4} on the $\nu_{\tau}$ magnetic moment 
generalizes straightforwardly to any light neutral fermion N which
can be produced in the reaction
$e^+ e^- \to N\bar{N}\gamma$: 

\begin{equation}
\mu_N < 4\cdot 10^{-6} \mu_B
\label{9}
\end{equation}

It is interesting to note that numerically close bound can be
obtained from measured at LEP invisible width of Z-boson.
Here the line of reasoning  goes as follows. If neutral fermion
is magnetically coupled to photon, it should couples to Z-boson
as well with strength damped by at most square of electroweak 
mixing angle sine. In this way invisible Z width get additional
contribution and experimental bound on the last leads to bound
on magnetic moment.

While this might still allow for a rather generous low momentum 
transfer (transition) magnetic moments through the above-proposed 
mechanism,this limit is comfortably
stronger than (\ref{1}).
Stronger bounds come from astrophysical data.
The bound from the white dwarf cooling \cite{5} $\mu_N <
3\cdot 10^{-11} \mu_B$ is much stronger.
Let us remind that this white dwarf bound is valid for masses of $N$
lower than $\omega_p /2 \approx 20$ KeV; since heavier particles can not be
produced in plasmon decay. 
As we have seen, the low-momentum transfer enhancement only
occurs for nearly-degenerate particles. 
So for a $\nu_e$ magnetic moment
generated through the proposed mechanism the  white dwarf bound is applicable
(as $m_N \approx m_{\nu_e} < 4$ eV); while for $\nu_{\mu}$ and
$\nu_{\tau}$ magnetic moments values up to the bound (\ref{9}) are still
allowed if they  (and hence $N$) happen to be heavier than 20 KeV.

In conclusion we demonstrate that a nontrivial $q^2$ dependence,
resulting in a strong low-momentum enhancement of a
neutrino magnetic form factor is possible: the bounds from $\nu
e$-scattering experiments at $q^2 \sim (1 {\rm MeV})^2 \div (100 {\rm
MeV})^2$ are not directly applicable to the neutrino magnetic moment
which governs the  neutrino behavior in a resonant cavity,
in a constant magnetic field, or even to radiative decays.

The construction leading to such enhancement (and presented more 
as a devil's advocate's argument) involves as a  necessary ingredient
a light neutral fermion $N$ with a nonzero magnetic moment, and the enhancement
has been shown to apply if a fine tuning of the masses leads to a nearly
divergent behaviour of the relevent loop integral in the infrared.

Even this possible enhancement cannot save the hope of observing induced
radiative transitions using a resonant cavity, due to the strong LEP limits
on the magnetic moment of any light particle, including the
hypothetical $N$.
Despite our efforts as devil's advocates, the  beautiful idea of 
serching for 
stimulated neutrino conversion in
a high-quality resonant cavity \cite{1}, \cite{2} stays less
sensitive than the experimental bounds
(\ref{2}), (\ref{3}), (\ref{8}), (\ref{9}) ( by at least two orders of
magnitude in the magnetic moments, that is four orders of magnitude
in the number of events).

As a complement to the present study, let us note recent proposal \cite{6} to look for neutrino magnetic
moment in reaction $\nu e \to \nu e \gamma$, where the interesting region
of small intermediate photon virtuality can be investigated.

We thank M.C.Gonzalez-Garcia for a discussion of experimental bounds on
neutral fermion magnetic moments, and JMF acknowledges 
at this occasion the support of the european network Flavourdynamics (chrx-ct93
-0132).

The investigations of J.-M.Fr\`ere, R.Nevzorov and M.Vysotsky were supported by
INTAS grant 94-2352; those of R.Nevzorov and
M.Vysotsky were supported by the grant RFBR-96-02-18010 as well.


\begin{thebibliography}{99}
\bibitem{1} S.Matsuki and K.Yamamoto, Phys. Lett. {\bf B289} (1992)
194.
\bibitem{2} M.C.Gonzalez-Garcia, F.Vannucci and J.Castromonte,
Phys. Lett. {\bf B373} (1996) 153.
\bibitem{3} Phys. Rev. {\bf D50} (1994) 1173, Review of Particle
Properties.
\bibitem{4}
%M.Acciarri et al., Phys.Lett. {\bf B346} (1995) 190.
H.Grotch and R.W.Robinett, Zeit. fur Physik {\bf C 39} (1988) 553.
\bibitem{5} S.I.Blinnikov, preprint ITEP-19 (1988); \\
S.I.Blinnikov, V.S.Imshennik and D.K.Nadyozhin, Sov. Sci. Rev. {\bf
E6}, 266 (1988).
\bibitem{6} J.Bernabeu, S.M.Bilenky, F.J.Botella and J.Segura,
Nucl.Phys. {\bf B246} (1994) 434.
\end{thebibliography}
\end{document}